\documentclass[UTF8,a4paper]{article}
\usepackage{amsmath}
\usepackage{bm}
\usepackage{graphicx}
\usepackage{subfigure}
\usepackage{ upgreek }
\usepackage{ stmaryrd }
\usepackage{ mathrsfs }
\usepackage{epstopdf}
\usepackage{ marvosym }
\usepackage{ tipa }
\usepackage{geometry}
\usepackage{ wasysym }
\usepackage{ulem}
\usepackage{indentfirst}
\usepackage{amsfonts,amssymb,dsfont}
\usepackage{accents}

\usepackage{setspace}
\usepackage{threeparttable}
\usepackage{amsmath,mathtools,amsthm}
\usepackage{array}
\usepackage{extarrows}
\usepackage{upgreek}

\makeatletter
\renewcommand*\env@matrix[1][\arraystretch]{%
	\edef\arraystretch{#1}%
	\hskip -\arraycolsep
	\let\@ifnextchar\new@ifnextchar
	\array{*\c@MaxMatrixCols c}}
\makeatother

\usepackage[pdfstartview=XYZ,
bookmarks=true,
colorlinks=true,
linkcolor=blue,
urlcolor=blue,
citecolor=blue,
pdftex,
bookmarks=true,
linktocpage=true, 
hyperindex=true
]{hyperref}
\usepackage{orcidlink}

	\title{Non-Hermitian effect to the ballistic transport and quantized Hall conductivity in 2H-MoS$_{2}$}

	\author{
		Chen-Huan Wu
	\orcidlink{0000-0003-1020-5977} 
		\thanks{chenhuanwu1@gmail.com},
		\ \ 
		Yida Li
		\orcidlink{0000-0002-5675-582X}
		\thanks{liyd3@sustech.edu.cn}
		\\
		School of Microelectronics, Southern University of Science and Technology, 518055 Shenzhen, China
	}
\begin{document}
	\maketitle
	\begin{small}
		\begin{abstract}
			By designing a multi-channel millimeter Hall measurement configuration,
			we realize the carrier-density (locally) controllable measurement on the transport property in 2H MoS$_{2}$.
			We observe a linearly increased Hall conductivity and exponentially decreased resistivity as the increase of dc current.
			The intrinsically large band gap does not exhibit too much effect on our measurement,
			as far as the magnetic field is above the critical value,
			which is $B=6$ T for 2H-MoS$_{2}$ in our experiment.
			Instead, the edge effect
			as a result of one-dimensional channels can be seen.
			This is different from the Corbino geometry which is widely applied on semiconductors, where the edges are absent.
			At room temperature,
			we observe that the emergent
			quantized quantum Hall plateaus are at the same values
			for both the two measurements,
			which implies that the quantized conductivity does not depends on the non-Hermitian interactions,
			but the number of partially/fully filled Landau levels,
			and this is in consistent with the previous theoretical works\cite{Siddiki}.
			At low-temperature limit,
			the Hall plateaus are only contributed by electrons above fermi energy,
			and in this case, the two measuremens exhibits stronger distinction,
			where we observe stronger fluctuations (on voltage, conductivity,
			and resistivity) at the currents between where there are
			Hall plateaus at higher temperature.
			\\

		\end{abstract}
		
		{\color{red} $Introduction$}
		The quantized Hall effect\cite{Siddiki} has a high reproducibility among a wide classes of materials,
		where the spatial distribution of effective carriers plays an important role in a Hall bar geometry
		which results in the incompressible strips with finite width.
		The quantization of conductivity as well as the distribution of incompressible densities then directly relate to the external magnetic field, temperature, and the materials energy band gap.
		Except the Hall conductivity,
		the quantization phenomenon also exhibited in the quantized
		exciton energies induced by strong Coulomb interaction\cite{Wang},
		exciton Hall effect\cite{Onga}.
		
		In this work,
		we propose and realize a Hall measurement with 
		incompressible strip-like local transport along the direction of applied dc current (the direction that is translational invariant),
		which can be realized in a wide class of wide band gap materials, like semiconductors.
		The confinement here originates from the distinct voltage applied in bulk and edge of the two-dimensioanl material.
		Our measurement exhibit similar features of the incompressible
		strips that realized in Hall bar base on the topological insulators\cite{Kendirlik},
		as well as the gate-relized one-dimensional channel\cite{Boddison}.
		Similar to Ref.\cite{Boddison},
		we artifically distingush the in-plane carrier transport into three sectors along the direction parallel to the drain and source eletrodes: the edge with higher conductance and bulk with higher resistance.

		Due to the intrinsic large bandgap and spin-valley locking,
		the valley Hall effect can be observed for monolayer or few layer 
		samples\cite{Wu}.
		Except the valley Hall effect,
		other phenomenons like the quantum spin Hall
		(QSH) effect as well as the topological insulting state can be generated by laser illuminating.
		Interestingly,
		the properties of topological insulator can sometimes
		appear in semiconductors under some external tunable conditions,
		like the atom-thin MoS$_{2}$ under laser\cite{Mine},
		or GaSa under tensile strain\cite{Zhao}.

		By designing a multi-channel millimeter Hall measurement configuration,
		we realize the carrier-density (locally) controllable measurement on the transport property in 2H MoS$_{2}$.
		Despite being a semiconductor with high resistance (up to $10^{7}-10^{9}$ $\Omega$ for the in-plane transport, but lower for interlayer transport),
		our experiment configuration allows the manufacture of electron confinement as well as
		the controllable inequilibrium distribution of the carrier.
		Through the turning of magnetic field and in-plane dc current,
		we can realize a transition from Hermitian transport (the quantized Hall conductivity with well defined Hall plateaus
		where each one can be expressed as an integrable eigenstate of a non-ergodic Hamiltonian, and corresponding to a restriction on the corresponding
		imcompressible carrier density),
		to the non-Hermitian transport (with stronger shart-range interacting effect and thus suppress the localization of the imcompressible carrier densities).
		Also,
		our result pave a way to studing the Hermitian to non-Hermitian (ballistic to diffusion) transition in low-dimensional materials
		(not just the topological insulators), and
		realizing the topological protection without the dependence on the device geometry.
		For the former, the integrability side promotes the ideal ballistic-like transport in a single-particle picture,
		and supply a Drude peak (zero frequency conductivity) for most metarials (expect the insulator).
		The key reason for the easily-realization of eletronic confinement as well as the distinct behavoir in a material
		is the coexistence of conducting edge (protected by the topological symmetries)
		and the insulating bulk.
		For strong enough topological effect,
		the edge can maintain the Hall quantization even if the bulk becomes compressible\cite{Kendirlik}.
		
		{\color{red} $Result\ and\ discussion$}
		By measuring the interlayer Hall conductivity at the edge of 2H-MoS$_{2}$,
		we found that when two channels of in-plane current are applied to 
		the edge and bulk parts, respectively,
		the quantization of Hall conductivity becomes more significant
		in the high current region, compares to the case without the bulk current.
		In consistent with experiment in Ref.\cite{Kendirlik,Boddison},
		our experiment result shows that,
		the conductance in the Hermitian (integrability) side exhibit a higher overall Hall conductivity
		compares to that in the non-Hermitian side.
		The reason is that, in the non-Hermitian side,
		the electron transport turns to diffusion-like,
		and the perturbation due to the interactions breaks the integrability.
		As a result,
		the stepwise conductivity will be squeezed downward or pulled toward the higher current.
		Both of them will cause the lowering of conductivity slope, and diluting the quantized chanracter.
		From the $\sigma_{xy}-I$ curve,
		we can clearly see the the pulling effect toward the rightside
		for the measurement without the bulk current,
		there are more mild plateau-to-plateau transition which correspond to the oscillations with larger periodicity
		and slower ramping up in the $V-I$ curve.
		
		In the integrable side, the ideal conductivity is supported by the fully occupied Landau levels and the
		Chern number directly determining the Hall conductance.
		Except the turn on/off of bulk current,
		we apply the same edge current for the two measurements,
		and thus there is a pulling effect in the $V-I$ curve, instead of a squeezing effect,
		i.e., the plateaus for the two measurements are at the same level of voltage,
		but appear at different current.
		A advantage of our measurement device is the large intrinsic resistance excluding the effects from traped carrier
		as well as the effect from original free carriers, and thus we are able to more precisely investigating the external-current-caused 
		interlayer transport.
		An evidence is emergent high-resistance peak for the measurement 1 near the $I=50 \mu A$,
		which corresponds to (ideally) a conducting edge and insulating bulk.
		The high resistance here, compares to measurement 2,
		indicates the strongest effect from imcompressible bulk and the vanishing correlation effect
		which is competive with the localization.
		While for measurement 2,
		the remained conductivity at $I=50\mu A$
		originates from the contribution of compressible bulk as well as the many-body
		interactions.
		Thus at this current,
		the effect of large intrinsic resistance overwhelming that of edge-bulk correlation.
		While at large current region, it is an opposite case.
		Also,
		we observe the edge burst effect from the left-side of each plateau in measurement 1,
		which corresponds to the minor enhancement of $R_{xy}$ at the large current.
		At low temperature,
		the measurement 1 realize a quasi 1D transport as the edge-edge scattering and edge-bulk scattering are
		suppressed by the artifically distincted carrier densities.

		At high temperature for the measurement 1,
		as shown in Fig.\ref{351}(a)-(b),
		the bulk current well prohibit the long-range scattering between two edges
		and the electron transport between top and bottom electrodes
		that contribute to the Hall current is squeezed from the bulk toward the edge
		while keeping the ballistic character.
		In thus case, the Hall conductivity can be viewed as a superposition of 
		the vertical electronic channels, and
		the coherence between the narrow electronic channels (under confinement)
		can nearly be ignored,
		and the quantized Hall plateaus are soly determined by the integrable eigenvalues
		where each one corresponds to a well-defined energy of noninteracting band.
		For measurement 2,
		as the non-Hermitian effect is suppressed by the finite temperature as well as the electron-electron interations,
		there is not emergent quasi-particle bands, and
		it requires a larger injected current to fill the same Landau level as the measurement 1 did.
		The case changes when lower the temperature
		to where the fermi wavelength is over the phase-breaking length,
		where the superposition of electronic channels is replaced by the statistical mixture of the identically
		distributed (mutually independent) ones.
		Thus in the presence of single-particle picture, which is validated by the dominating imcompressibility and the probidented charge redistribution,
		he strongest localization (shortest localization length) appears in teh center of each plateau,
		while the highest resistivity appears in the right-side of each plateau
		(opposite to the edge burst for conductivity which is in the left side of plateau).

		At low-temperature,
		as shown in Fig.\ref{351}(c)-(d),
		for measurement 1,
		the bulk current simulate an insulating effect by enhance the imcompressibility (e.g., through the backscattering),
		as a result,
		the interlayer electron transport will be squeezed toward the edge
		as experimentally observed in valley Hall effect in gapped graphene.
		That explain the remained step-like conductivity at high injected current.
		In the mean time, as a result of long-range scattering and invalidated topological confinement,
		the edge channels are heavily localized with shortened mean free path and
		the weak conductivity mainly comes from the hopping between localized states.
		While for measurement 1,
		the remained high conductivity comes from the entanglement type interaction
		whose weak effect can be seem from the relatively confinement on the narrowed channels
		along the direction of injected current, and the translational invariance in this direction
		as a nonlocal symmetry could be produced by the elimination on the diffusions on the direction perpendicular to injected current.
		As a result,
		the quasi-one-dimensional ballistic transport will modified fraction (according to the distance to the
		edge or that away from the bulk center).
		Unlike the case of integrability where the superposition of multi-momenta (which is good quantum number)
		reported in Ref.\cite{Li},
		the non-Hermitian effect emergent as a result of nonreciprocity.
		The broken integrability can be seem from the conductivities in Fig.\ref{351}(c)-(d),
		where the single plateau for measuremnet 1 in divided into there steeper pieces,
		and that means the increase of conductivity is nomore according to the filling of original levels,
		but a new stable level distribution which can be regarded as the distribution at thermodynamic equilibrium.
		In this case, the effect of charge compressibility is more obviously at measurement 1 compares to 
		that at measurement 2,
		as can be seem from the higher fraction (number of Landau level under fermi level) of measurement 2
		in Fig.\ref{351}(c)-(d),
		despite the strict condition in the strength of magnetic field to observe such effect
		(compares to the same results at magnetic field $B=7,9$ T).
		Also, this is in contrast to the case of higher temperature,
		where
		measurement 2 exhibit stronger compressibility and thus will lower fraction
		as shown in Fig.\ref{351}(a)-(b).

	At room temperature,
		we observe the higher Hall conductivity in a semiconductor in measurement 1, where the source-drain current was applied to provide a homogeneous carrier density distribution along the direction of edge
		and meanwhile a current of opposite direction was applied within the bulk
		to realizing a electronic confinement (like the lateral confinement in quantum well with the increased mobility and lowered dimension),
		compares to measurement 2 (without the bulk current).
		This can be explained by the artifically-relaized topological effect where
		the bulk is relatively imcompressible compares to the edge
		and bulit a obstruction for the diffusion-like transport in the edge.
		At low-temperature limit, where the obstruction effect in measurement 1 disappear,
		we observed higher conductivity and stronger fluctuation in measurement 2.
		Our explaination is base on the non-Hermitian effect which enfore the confinement
		(not the coherence-type) and
		results in a continuously varied charge distribution from center of bulk to the edge,
		and the ergodic character can be seem from the experimental results at $T=2$ K for different magnetic field,
		i.e., the narrowed conductivity steps and enhanced fluctuation between adjacent plateaus
		(teh changed Hall fraction factor strongly implies the emergent low-energy band built by the non-Hermitian interactions).
		Note that the non-Hermitian interaction here is of the short range type such that there
		is ergodic and entanglement characters between arbitarily two narrow channels (along the direction of current).
		Unlike the similar transport in thin system, the intrinsically large band gap and the absence of 
		edge state
		penetration into the bulk,
		the larger conductivity in measurement 2 can be almostly regarded as originates from the non-Hermiticity,
		where the raised quasiparticle (quasi-one-dimensional) bands whose energy distributions become
		asymptotically Gaussian by virtue of the central limit theorem.
		While the measuremnet 1 with opposite current in the bulk
		breaks the ergodicity by generating a topological-protected edge (from the short-range scatterings)
		despite is cannot obstract the long-range scattering (between two edges) anymore.
		Similar experimental results is reported in Res.\cite{Li,zhu}
		base on the thin topological insulators.
		In these works,
		the persistive high conductivity at low-temperature was explained by the 
		remnant charge inhomogeneity which leads to the hopping conductivity between localized states,
		and the ideal conductivity relys on the penetration
		length and thus it is still of the integrable case (i.e., the penetration
		length is determined by the electron velocity/mass of the initial edge state.
		Specifically, 
		a strong disorder case was mentioned in Ref.\cite{Li},
		where penetration length of different edge states converges to
		a same value at strong disorder.
This indeed meets our expectation considering the non-Hermitian effect:
		As there are large enough amount of random edges
		each of which penetrate into the bulk in a certain distance,
		there overall effect (including the lowered energy gap and weakened Coulomb screening\cite{wu,pol})
		will cause a largest wavefunction overlap in a position near the edge.
		Such initial (before the penetration) edge condition-independent 
		effect can be explained well by the statistical mixture of their wavefunctions.
		Beside the contribution from edge states extend into the bulk,
		the insulating state in edge is due to the defeat-induced back scattering in thin materials,
		and due to the diffusion-induced broken quasi-one-dimensional topological edge state
		in this work.
		
		\begin{figure}
			\centering
			\includegraphics[width=0.8\linewidth]{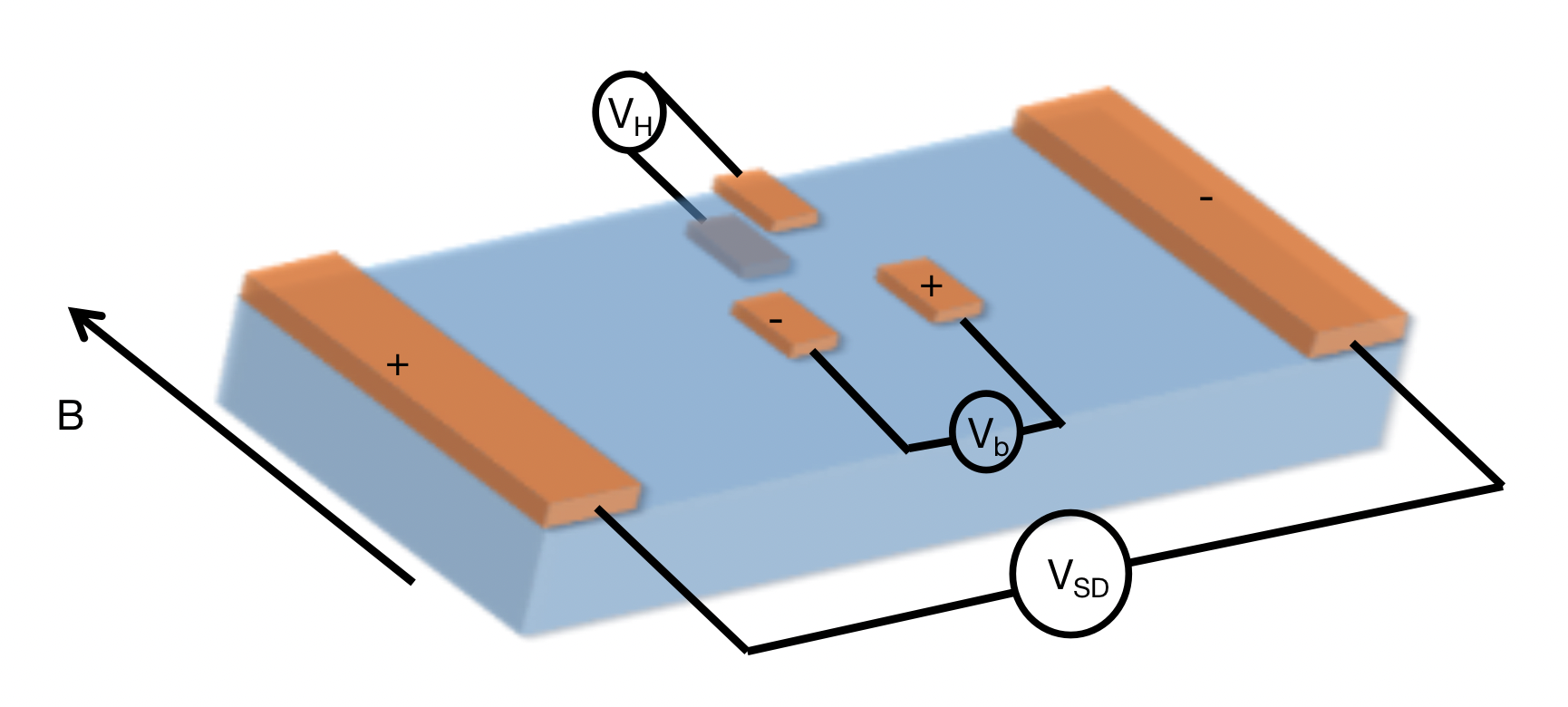}
			\caption{
					Sketch for the experimental set-up base on 2H-MoS$_{2}$.
					There are two long source-drain contacts ($V_{SD}$) at the left and right sides, and two bulk contact ($V_{b}$).
					The interlayer Hall voltage is measured through the top and bottom contacts near the edge.
					The contacts in bulk simulate the insulating states due to the localization effect of magnetic field,
					where the imcompressible densities 
					has the fermi level falls into the Landau level gap,
					and the plateau intervals is much more sensitive to the magnetic field and the electric current.
					Next we refer the measurement with $V_{b}\neq 0,V_{SD}\neq 0$ by measurement 1, and $V_{b}=0,V_{SD}\neq 0$ by measurement 2.
			}
			\label{350}
		\end{figure}

		\begin{figure}
			\centering
			\includegraphics[width=0.9\linewidth]{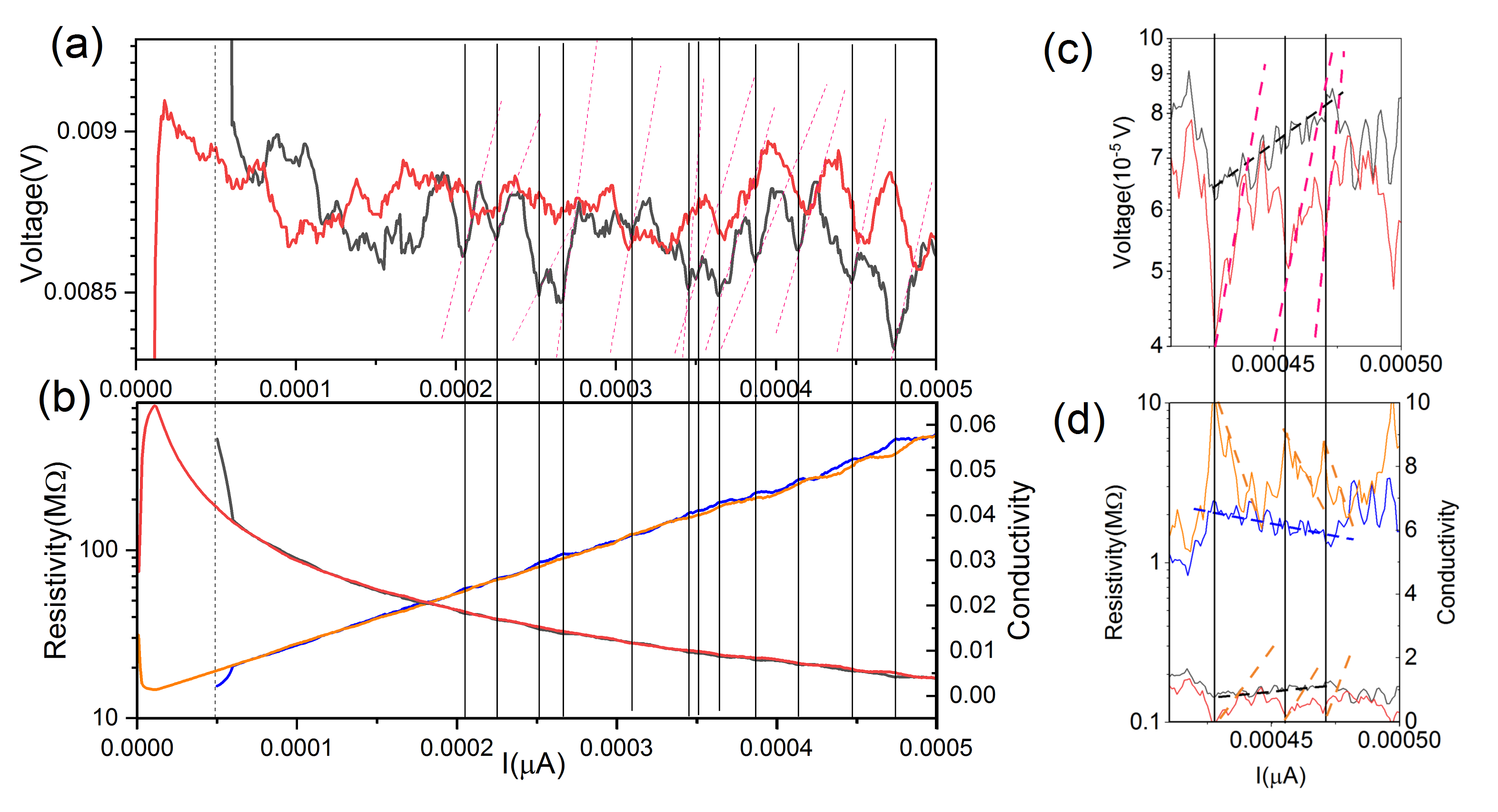}
			\caption{
					Voltage, resistivity, and conductivity under Hall measurement with magnetic field $B=8$ T at room temperature.
					For conductivity, the vertical axis is in unit of 0.0257692 $\frac{e^{2}}{h}$.
					The red and orange (black and blue)
					curves corresponds to measurement 1
					(measurement 2).
					The vertical black solid line indicates the positions of quantized Hall conductivity,
					and the dashed purple line indicates the slope of ramping voltage in each piece of quantized conductance.
					Despite no shown,
we found that at higher magnetic field ($>8$ T),
					the quantized conductivity appear at higher electric current.
					Thus the quantized conductivity is hard to realized at strong magnetic field where the carrier density trends to compressible.
					However, our experiments show that the Hall conductivity cannot longer be observed for magnetic field below $B=6$ T,
					due to the extremely large intrinsic resistance.
					The adjustable critical current where the quantization appear also reflects the controllable electronic confinement
					(to quasi-one-dimensional) along the direction of applied current.
(c)
					The same with (a)-(b) but at low-temperature $T=2$ K.
The Hermitian-type step-like conductivity is more obvious but with 
more drastic fluctuations at the low-temperature,
which is mainly due to the transport dominated by the the non-Hermitian effect.
The Fermi wavelength (Broglie wavelength) becomes very large compares to the width of the bulk imcompressible density at low-temperature, and the carrier density becomes lower,
					then the fermi distribution under the length scale of Fermi wavelength
					does not affected by the obstruct effect from the bulk 
					and thus scattering between left and right edges are allowed.Importantly,
					the measurement 2 exhibits a higher overall conductivity than the measurement 1,
					which is distinct from the case of room temperature shown in (a)-(b).
					This is due to the consumed contributions from the carriers that participate the scattering between two compressible edge states.
					As indicated by the dashed lines,
around the current up to $0.00045\ \mu A$,
					the non-Hermitian interactions change the original single plateau (blue dashed line in (d)) into three plateaus (orange dashed line in (d);
with shorter lasting time and larger slope) which is due to the effect of quasiparticles generated by the non-Hermitian interactions instead of the noninteracting band.
			}
			\label{351}
		\end{figure}

		\begin{figure}
			\centering
			\includegraphics[width=0.9\linewidth]{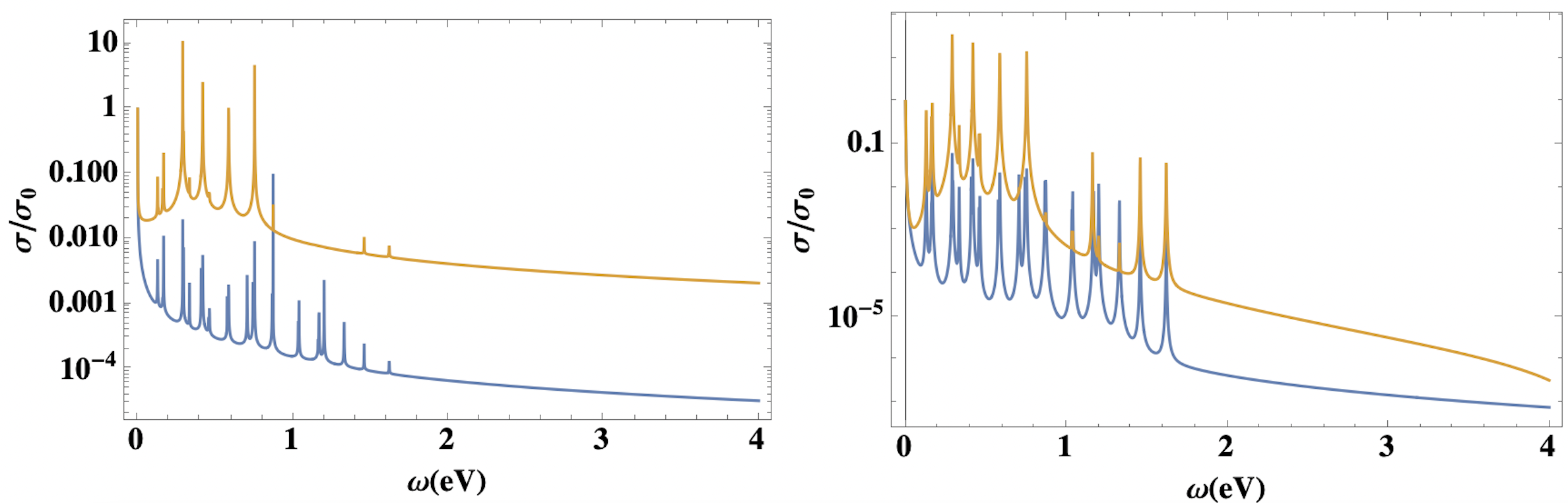}
			\caption{
				(Left)
					The corresponding optical conductivity including both the interband and intraband (Drude) transitions.
					The delta function of broadened form is used here.
					(Right) The same with (b) but with delta function expressed by the advanced/retarded Green function.
			}
			\label{353}
		\end{figure}

		{\color{red} $Numerics$}
		As the transition metal dichalcogenides (TMDs) in semiconductor phase have wide gap and the available conduction bands for the Hall conductance 
		are of high-energy,
		we consider only the frequency-dependence of the conductivity
		and ignore the momentum-dependence,
		which is equavalents to the multi-flat-band model.
		Firstly we consider the case of optical conductivity,
		i.e., the finite external frequencies 
		allow the interband transition far above the fermi surface
		which can be written as
		\begin{equation} 
			\begin{aligned}
				\sigma_{H}(\omega)
				=\sum_{ij}
				\delta(\omega-(\varepsilon_{i}-\varepsilon_{j}))\frac{{\bf U}^{(1)}_{ij}{\bf U}^{(2)}_{ji}}
				{\varepsilon_{i}-\varepsilon_{j}}(1-\delta_{ij})
				-\sum_{i}\sum_{k=1,2}
				\delta(\omega)\delta(-\varepsilon_{i})
				|{\bf U}^{(k)}_{ii}|^2,
				\\
				\sigma_{nH}(\omega)
				=\sum_{ij}
				\delta(\omega-(\varepsilon_{i}-\varepsilon_{j}))\frac{{\bf U}^{(1)}_{ii}{\bf U}^{(2)}_{jj}}
				{\varepsilon_{i}-\varepsilon_{j}}(1-\delta_{ij})
				-\sum_{i}\sum_{k=1,2}
				\delta(\omega)\delta(-\varepsilon_{i})
				|{\bf U}^{(k)}_{ii}|^2,
			\end{aligned}
		\end{equation}
		where 
		\begin{equation} 
			\begin{aligned}
				{\bf U}^{(k)}_{ij}=\langle\varepsilon_{i}|{\bf J}_{k}|\varepsilon_{j}\rangle,
			\end{aligned}
		\end{equation}
		where $k=1,2$ correspond to the two subsystems in the current bipartite system.
		For $e^{iHt}{\bf J}e^{-iHt}$ exhibit entanglement among arbitary time in the evolutionary spectrum,
		we have ${\bf U}_{ij}{\bf U}_{ij}=
		{\bf U}_{ii}{\bf U}_{jj}$,
		while for that without the entanglement
		(but only the coherence),
		we have 
		${\bf U}_{ij}{\bf U}_{ij}<
		{\bf U}_{ii}{\bf U}_{jj}$.
		These are the two cases of Cauchy inequality.
		Only for those observables ${\bf J}$ which satisfies
		${\bf U}_{ij}{\bf U}_{ij}<
		{\bf U}_{ii}{\bf U}_{jj}$
		contributes to the distinct conductance between
		the Hermitian $\sigma_{H}(\omega)$
		and non-Hermitian $\sigma_{nH}(\omega)$.

		\begin{figure}
			\centering
			\includegraphics[width=0.8\linewidth]{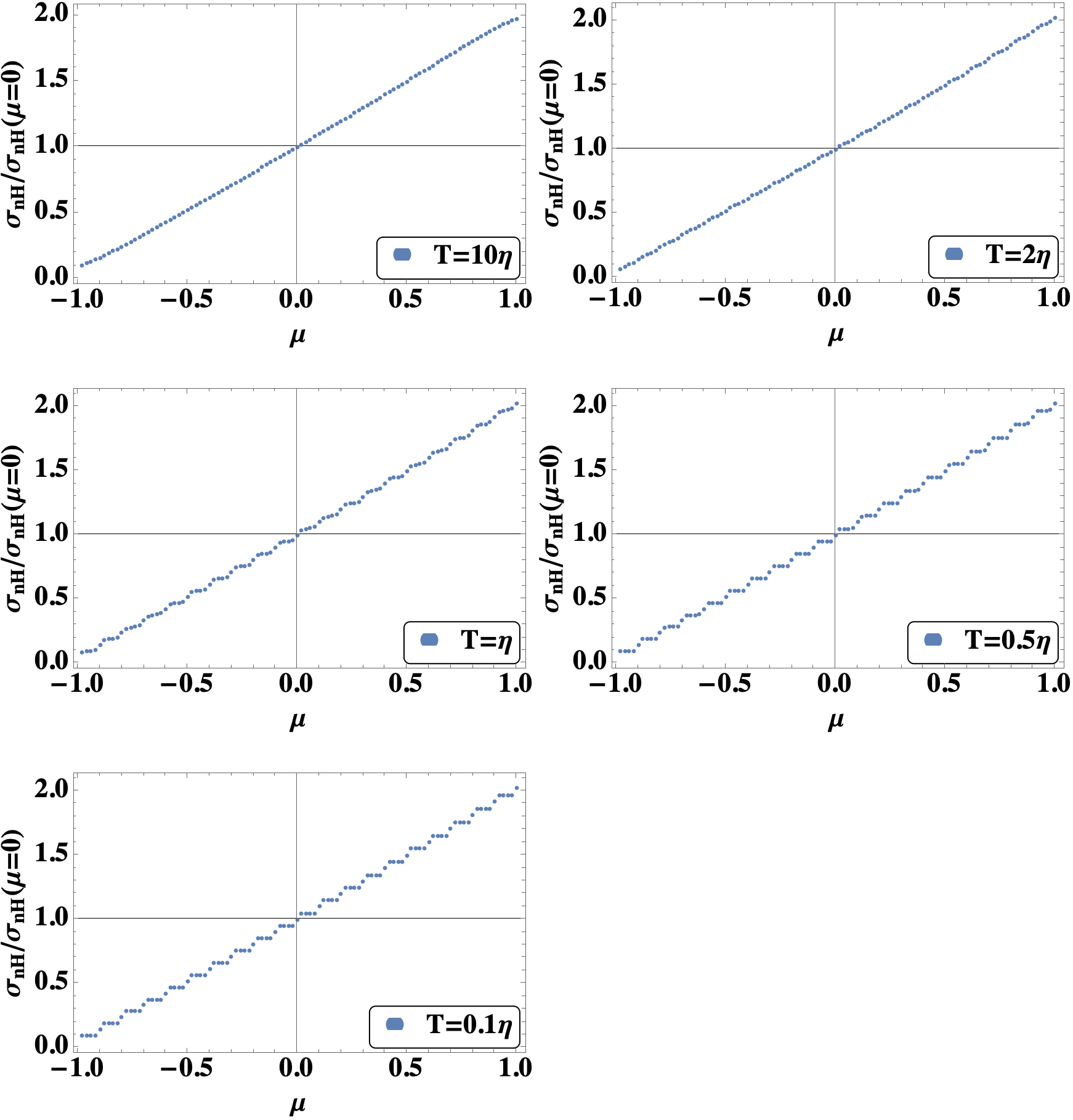}
			\caption{
					Non-Hermitian contribution to the Hall conductivity, as a function of chemical potential, at different effective temperature.
					Note that here the emergent step-like character should not be confused with the suppressed charge imcompressiblity at low-temperature (Fig.\ref{351}(c)-(d))
which is simply due to the enhanced noise.				
			}
			\label{1118}
		\end{figure}

		While for the case of dc conductivity at zero frequency,
		the conductivity can be expressed in terms of Kubo Bastin formula\cite{Garcia,Crepieux}
		\begin{equation} 
			\begin{aligned}
				\sigma_{H}(\mu)
				=\int d\varepsilon
				f(\varepsilon-\mu)
				\sum_{ij}\left[
				{\bf U}^{(1)}_{ij}\frac{\partial G^{-}_{i}}{\partial\varepsilon}
				{\bf U}^{(2)}_{ji}\delta(\varepsilon-\varepsilon_{j})
				-{\bf U}^{(1)}_{ij}\delta(\varepsilon-\varepsilon_{i})
				{\bf U}^{(2)}_{ji}\frac{\partial G^{+}_{j}}{\partial\varepsilon}
				\right],
			\end{aligned}
		\end{equation}
		where the Green function reads
		$G^{\pm}_{i}=(\varepsilon-\varepsilon_{i}\pm i\eta)^{-1}$,
		with $\eta$ a positive infinitely small quantity.
		As we select the system containing finite number of (nearly flat)
		bands whose spectrum is restricted within $[-1,1]$,
		we only need to perform the intergral of $\varepsilon$ over this range.
		Numerically, we shown in Fig.\ref{1118} the non-Hermitian effect-induced contribution to the Hall conductivity,
		where the two tensors ${\bf J}_{1}$ and ${\bf J}_{2}$
		are essential in forming the identically-distributed random
		Gaussian variables.
		which are the gredient in forming the non-Hermitian operator $\mathcal{H}_{nH}$.
		the numerical result shows that the step-like quantized character quickly vanishes in the
		zero-temperature limit,
		where the temperature here should be referred to the effect temperature as we mentioned above
		(different to the ones in experimental result in Fig.(\ref{351})).
		
			The quantized conductivity in Fig.\ref{1118} is a statistical result of the carrier transport where the non-Hermtian 
					(entanglement-like) interactions weaken the restriction that only the electrons 
					at fermi level contribute to current
					(and consequently the interactions are dominated by the long-range type scattering above the fermi wavelength).
This can be explained in perspective of statistic.
					The level repulsion in an ergodic system with low disorder
					exhibit level repulsion between the continuously distributed by no degenerated levels,
					where the spatial effect as well as the position of charge neutrality point becomes less important.
					That means, part of the electrons away from fermi level may also contributes to the current
					due to the entanglement interaction (which is of the short-range type).
					This can be observed by the homogeneously spaced plateaus,
					which is distinct to the quantized conductivity contributed by the narrow channels at high temperature
					where the shape of each plateau is determined by the corresponding integrable eigenstates.
					For the latter case,
					the plateaus of Hall conductivity are usually not equally spaced and have different height and length\cite{Garcia}.
					Also, the above discussion can be used to explain the suppressed insulating phase
					at low-temperature limit\cite{zhu},
					where the
					remaining charge imcompressibility
					and the residue charge density down to low-energy at low-temperature is observed.
					Ref.\cite{zhu} explain the low resistivity and long tail of saturated charge density in gapped graphene
					by the zigzag states with a long penetration length (corresponds to long-range scattering),
					which results in a quasi-one-dimensional impurity band 
					inside the bulk and a supercurrent narrowed near the edge
					that is dominated by the hopping-type conductivity.
					Such that the bulk become localized and imcompressible while the edge is delocalized.
					While in this work,
					we explain the persistive conductivity at the edge in perspective of non-Hermitian contribution,
					where the relatively higher conductivity in measurement 2 is due to the short-range
					entanglement-type interaction which the effect of level repulsion can be realized 
					by the identically distributed narrow channels along the direction of the applied electric current,
					and the role of spatial gradient (from center of bulk to edge) becomes less important.

		{\color{red} $Conclusion$}
		We observe a linearly increased Hall conductivity and exponentially decreased resistivity as the increase of dc current.
		The intrinsically large band gap does not exhibit too much effect on our measurement,
		as far as the magnetic field is above the critical value,
		which is $B=6$ T for 2H-MoS$_{2}$.
		Instead, the edge effect
		which emerge as a result of one-dimensional channels.
		This is different from the Corbino geometry which is widely applied on semiconductors, where the edges are absent.
		At room temperature,
		we observe that the emergent
		quantized quantum Hall plateaus are at the same value
		for both the two measurements,
		which implies that the quantized conductivity does not depends on the non-Hermitian interactions,
		but the number of partially filled Landau levels,
		and this is in consistent with the previous theoretical works\cite{Siddiki}.
		At low-temperature limit,
		the Hall plateaus are contributed
		by the electrons at fermi energy,
and the narrowed imcompressible region leads to longer-range scattering as well as more significant noise.
		In this case, the two measuremens exhibits stronger distinction,
		where we observe stronger fluctuations (of voltage, conductivity,
		and resistivity) at the currents between where there are
		Hall plateaus at higher temperature.
		The homogeneous charge densities and the nonintegracting edge states are only available at high temperature.

		Our study as well as the measurement setup can be further extended to the semiconductor 
		heterostructures,
		where a
		nearly decoupled potential steps has been observed in WSe$_{2}$/MoS$_{2}$ multi-layers\cite{Deb} which
		indicating highly confined interfacial electric felds 
		such that there is a ladder-type ferroelectricity
		and potentially related to the recently found 
		fractional quantum ferroelectricity\cite{Ji}.

		\renewcommand\refname{References}

		\clearpage
		
		{\color{red} $Appendix:\ Numerical\ calculation\ on\ Hall\ conductivity$}

		The Hall conductivity at zero frequency
		can be expressed in terms of Kubo Bastin formula\cite{Garcia,Crepieux}
		\begin{equation} 
			\begin{aligned}
				\sigma_{H}(\mu)
				=\int d\varepsilon
				f(\varepsilon-\mu)
				\sum_{ij}\left[
				{\bf U}^{(1)}_{ij}\frac{\partial G^{-}_{i}}{\partial\varepsilon}
				{\bf U}^{(2)}_{ji}\delta(\varepsilon-\varepsilon_{j})
				-{\bf U}^{(1)}_{ij}\delta(\varepsilon-\varepsilon_{i})
				{\bf U}^{(2)}_{ji}\frac{\partial G^{+}_{j}}{\partial\varepsilon}
				\right],
			\end{aligned}
		\end{equation}
		where the Green function reads
		$G^{\pm}_{i}=(\varepsilon-\varepsilon_{i}\pm i\eta)^{-1}$,
		with $\eta$ a positive infinitely small quantity.
		As we select the system containing finite number of (nearly flat)
		bands whose spectrum is restricted within $[-1,1]$,
		we only need to perform the intergral of $\varepsilon$ over this range.
		We use the formula
		\begin{equation} 
			\begin{aligned}
				\delta(\varepsilon-\varepsilon_{i})
				=\frac{1}{2i\pi}
				\left(\frac{1}{\varepsilon-\varepsilon_{i}-i\eta}
				-\frac{1}{\varepsilon-\varepsilon_{i}+i\eta}
				\right)
				=\frac{\eta}{\pi}
				\frac{1}{(\varepsilon-\varepsilon_{i})^2-(i\eta)^2},
			\end{aligned}
		\end{equation}
		thus we have 
		\begin{equation} 
			\begin{aligned}		\label{1114}
				\frac{\partial}{\partial \varepsilon}
				\delta(\varepsilon-\varepsilon_{i})
				=-\frac{\partial}{\partial \varepsilon}
				\delta(\varepsilon-\varepsilon_{i})
				=\delta(\varepsilon-\varepsilon_{i})
				\frac{-1}{\varepsilon-\varepsilon_{i}},
			\end{aligned}
		\end{equation}
		by using derivative
		\begin{equation} 
			\begin{aligned}
				&
				\lim_{\eta\rightarrow 0^{+}}
				\frac{\partial}{\partial \varepsilon}
				G^{\pm}_{i}
				=\lim_{\eta\rightarrow 0^{+}}
				\frac{\partial}{\partial \varepsilon}
				\frac{1}{\varepsilon-\varepsilon_{i}\pm i\eta}
				=\lim_{\eta\rightarrow 0^{+}}
				\frac{-1}{(\varepsilon-\varepsilon_{i})(\varepsilon-\varepsilon_{i}\pm i\eta)}.
			\end{aligned}
		\end{equation}
		Considering the Hermitian contribution from the real energy $\varepsilon$ from the Hermitian part,
		we have 
		\begin{equation} 
			\begin{aligned}
				&
				\lim_{\eta\rightarrow 0^{+}}
				\frac{\partial}{\partial \varepsilon}{\rm ln}
				\frac{1}{\varepsilon-\varepsilon_{i}\pm i\eta}
				=
				\frac{-1}{\varepsilon-\varepsilon_{i}}.
			\end{aligned}
		\end{equation}
		For the selected eigenvalue of integrable eigenstate reaches the maximal ($\varepsilon_{i}=1$),
		the variational behavior of $G_{i}^{\pm}(\varepsilon_{i}=1)$ follows the polylogarithm function,
		${\rm ln}
		\lim_{\eta\rightarrow 0^{+}}
		\frac{d}{d\varepsilon}
		{\rm ln}G_{i}^{\pm}(\varepsilon_{i}=1)
		={\rm Li}_{1}(\varepsilon)$,
		such that,
		$\frac{\partial}{\partial \varepsilon}
		{\rm Li}_{1}(\varepsilon)=\frac{1}{\varepsilon}
		{\rm Li}_{0}(\varepsilon)$,
		$\frac{d}{d\varepsilon}
		\frac{1}{\varepsilon}
		{\rm Li}_{0}(\varepsilon)=\frac{\partial}{\partial \varepsilon}
		{\rm  Li}_{0}(\varepsilon)=
		(\frac{{\rm Li}_{0}(\varepsilon)}{\varepsilon})^2$.

		\begin{equation} 
			\begin{aligned}
				&
				{\rm ln}\frac{-G^{+}_{i}}{G^{-}_{i}}=
				i\pi+{\rm ln}\frac{G^{+}_{i}}{G_{i}^{-}}=i\pi{\rm sgn}(\varepsilon-\varepsilon_{i}),\\
				&
				\frac{\partial}{\partial \varepsilon}{\rm ln}\frac{G^{+}_{i}}{G_{i}^{-}}=
				\frac{\partial}{\partial \varepsilon} i\pi{\rm sgn}(\varepsilon-\varepsilon_{i})=
				2 i\pi\delta(\varepsilon-\varepsilon_{i})=G^{-}_{i}-G^{+}_{i}
				=2i\eta G^{+}_{i}G^{-}_{i},\\
				&
				\frac{\partial}{\partial\varepsilon}{\rm ln} G^{+}_{i}G_{i}^{-}
				=
				\frac{\partial}{\partial\varepsilon}{\rm ln} \left(\frac{\pi}{\eta}\delta(\varepsilon-\varepsilon_{i})\right),
			\end{aligned}
		\end{equation}
		Combine with Eq.(\ref{1114}),
		we have
		\begin{equation} 
			\begin{aligned}
				&
				\frac{\partial_{\varepsilon}\delta(\varepsilon-\varepsilon_{i})}{\delta(\varepsilon-\varepsilon_{i})}
				=\partial_{\varepsilon}{\rm ln}\delta(\varepsilon-\varepsilon_{i})
				=\frac{-1}{\varepsilon-\varepsilon_{i}},\\
				&
				\frac{\partial_{\varepsilon}\delta(\varepsilon-\varepsilon_{i})}{\delta(\varepsilon-\varepsilon_{i})}
				=\frac{\partial_{\varepsilon}G^{-}_{i}-\partial_{\varepsilon}G^{+}_{i}}{G^{-}_{i}-G^{+}_{i}}
				=\frac{G^{-}_{i}\partial_{\varepsilon}{\rm ln}G^{-}_{i}
					-G^{+}_{i}\partial_{\varepsilon}{\rm ln}G^{+}_{i}}{G^{-}_{i}-G^{+}_{i}}
			\end{aligned}
		\end{equation}
		where we note that
		$\partial_{\varepsilon}G^{\pm}_{i}=\frac{1}{2}\left(\pm
		\frac{\partial}{\partial\varepsilon}{\rm ln} \frac{G^{+}_{i}}{G_{i}^{-}}+
		\frac{\partial}{\partial\varepsilon}{\rm ln} G^{+}_{i}G_{i}^{-}\right)$.
		Then in the limit of
		$\frac{\partial}{\partial\varepsilon}{\rm ln} \left(\frac{\pi}{\eta}\delta(\varepsilon-\varepsilon_{i})\right)
		\rightarrow
		\frac{\partial}{\partial\varepsilon}{\rm ln}\delta(\varepsilon-\varepsilon_{i})$,
		we have
		$\frac{\partial}{\partial\varepsilon}{\rm ln}\delta(\varepsilon-\varepsilon_{i})
		=\frac{\partial}{\partial\varepsilon}{\rm ln}\left(\frac{\pi}{\eta}
		\delta(\varepsilon-\varepsilon_{i})\right)
		\approx -G_{i}^{-}-G_{i}^{+}
		=\frac{-2(\varepsilon-\varepsilon_{i})}
		{(\varepsilon-\varepsilon_{i})^2-(i\eta)^2}$,
		thus $\eta^2=(\varepsilon-\varepsilon_{i})^2$,
		where correspondingly,
		\begin{equation} 
			\begin{aligned}
				&
				\partial_{\varepsilon}\frac{G^{+}_{i}}{G^{-}_{i}}=
				G^{-}_{i}-G^{+}_{i}=\frac{i}{\eta},\\
				&
				\delta(\varepsilon-\varepsilon_{i})=\delta(\eta)=\frac{1}{2\pi\eta},\\
				&
				\partial_{\varepsilon}{\rm ln}G^{+}_{i}G^{-}_{i}=
				\partial_{\varepsilon}{\rm ln}\frac{\pi}{\eta}\delta(\varepsilon-\varepsilon_{i})
				=\frac{-1}{\varepsilon-\varepsilon_{i}}
				\frac{2\eta^2}{(\varepsilon-\varepsilon_{i})^2-(i\eta)^2}
				=\frac{-1}{\varepsilon-\varepsilon_{i}},\\
				&
				\partial_{\varepsilon}{\rm ln}G^{\pm}_{i}=-G^{\pm}_{i},\\
				&
				(G^{\pm}_{i})^2=\frac{1}{\pm 2(\varepsilon-\varepsilon_{i})i\eta}.
			\end{aligned}
		\end{equation}
		In the mean time,
		for the case of complex energy (like the non-Hermitian system),
		we have
		\begin{equation} 
			\begin{aligned}
				\frac{\partial}{\partial \varepsilon^{*}}
				\partial_{\varepsilon}{\rm ln}\delta(\varepsilon-\varepsilon_{i})=
				\frac{\partial}{\partial \varepsilon^{*}}\frac{-1}{\varepsilon-\varepsilon_{i}}
				=
				\left(\frac{\partial}{\partial \varepsilon}\frac{-1}{\varepsilon-\varepsilon_{i}}\right)
				\left(\frac{\partial}{\partial \varepsilon^{*}}(\varepsilon-\varepsilon_{i})\right)
				=
				\pi \delta(\varepsilon-\varepsilon_{i}),
			\end{aligned}
		\end{equation}
		which leads to
		$\eta=2\frac{\partial}{\partial \varepsilon^{*}}(\varepsilon-\varepsilon_{i})$. 
		Note that the delta function containing such a complex variable
		satisfies
		$\int\int \frac{d\varepsilon \wedge
			d\varepsilon^{*}}{-2i}\delta(\varepsilon-\varepsilon_{i})f(\varepsilon)
		=\frac{-1}{\pi}\int\int
		\frac{f(\varepsilon)}{\varepsilon-\varepsilon_{i}}
		\frac{d\varepsilon}{-2i}
		=f(\varepsilon_{i})$,
		for holomorphic function $f(\varepsilon)$.
		At low-temperature case if we cast the 
		energy-difference term $(\varepsilon-\varepsilon_{i})$ into the integral over a 
		Heaviside step function,
		$\varepsilon-\varepsilon_{i}=\int d\varepsilon'
		f(\varepsilon,\varepsilon')$,
		such that
		$\partial_{\varepsilon}(\varepsilon-\varepsilon_{i})
		=\int d\varepsilon' \partial_{\varepsilon} f(\varepsilon,\varepsilon')
		=\int d\varepsilon' \delta(\varepsilon,\varepsilon')=1$,
		we can further obtain
		$f(\varepsilon_{i})=\frac{{\rm ln}(\varepsilon-\varepsilon_{i})}{2i\pi}$.

		For the non-Hermitian case,
		the Hall conductivity follows the form
		\begin{equation} 
			\begin{aligned}
				\sigma_{nH}(\mu)
				=\int d\varepsilon
				f(\varepsilon-\mu)
				\sum_{ij}\left[
				{\bf U}^{(1)}_{ij}\frac{\partial G^{-}_{i}({\bf J})}{\partial\varepsilon}
				{\bf U}^{(2)}_{ji}\delta(\varepsilon-\varepsilon_{j}
				-\varepsilon_{nH})
				-{\bf U}^{(1)}_{ij}\delta(\varepsilon-\varepsilon_{i}
				-\varepsilon_{nH})
				{\bf U}^{(2)}_{ji}\frac{\partial G^{+}_{j}({\bf J})}{\partial \varepsilon}
				\right],
			\end{aligned}
		\end{equation}
		where $G^{\pm}_{i}({\bf J})
		=(\varepsilon-\varepsilon_{i}\pm i\eta
		-\varepsilon_{nH})^{-1}$ is the interacting Green function.
		$\varepsilon_{nH}=
		-{\rm Tr}
		[|\varepsilon_{i}\rangle\langle\varepsilon_{i}|
		{\bf J}_{1}]
		-{\rm Tr}
		[|\varepsilon_{i}\rangle\langle\varepsilon_{i}|
		{\bf J}_{2}]$.
		The traces
		${\rm Tr}
		[|\varepsilon_{i}\rangle\langle\varepsilon_{i}|
		{\bf J}]$ ($i=1,\cdots,8$) for the product between initial density operator and the tensor operator ${\bf J}$,
		equals to the eigenvalue of non-Hermitian observable
		(whose temporal fluctuation satisfies the ETH ansatz).

		For the fermi distribution at low-temperature limit,
		\begin{equation} 
			\begin{aligned}
				\lim_{T\rightarrow 0}\partial_{\varepsilon}f(\varepsilon-\mu)=-\delta(\varepsilon-\mu)
				=-\delta(\varepsilon-2\mu)
				\bigg|\frac{\partial (\varepsilon-\mu)}{\partial \varepsilon^{*}}\bigg|
				_{\varepsilon^{*}=\mu^{*}}\bigg|^{-1},
			\end{aligned}
		\end{equation}
		then since 
		$\frac{\partial}{\partial \varepsilon^{*}}(\varepsilon-\varepsilon_{i})
		=\frac{\eta}{2}$,
		we obtain
		$|\frac{\eta}{2}|=\frac{\delta(\varepsilon-2\mu)}{\delta(\varepsilon-\mu)}$.
		
		Since the interaction-induced non-Hermitian transport as well as the emergent entanglement entropy 
		during the thermalization,
		we can consider the non-Hermitian effect as the thermal entropy base on an effective inverse temperature
		$\beta(=\frac{1}{T})$,
		\begin{equation} 
			\begin{aligned}\label{1117}
				\mathcal{S}={\rm ln}\mathcal{Z}+T\frac{\partial}{\partial T}{\rm ln}\mathcal{Z}
				={\rm ln}\mathcal{Z}+\frac{1}{T}
				\frac{{\rm Tr}[\mathcal{O}\mathcal{H}_{lo}]}{\mathcal{Z}},
			\end{aligned}
		\end{equation}
		where $\mathcal{Z}={\rm Tr}[\mathcal{O}]$ is the partition function, i.e., the trace over
		an non-Hermitian observable $\mathcal{O}$ which is a thermalized density operator.
		Inspired by the von Neumann entropy,
		we replace the $\beta$ by ${\rm ln}\beta$ during the analysing of thermodynamics,
		\begin{equation} 
			\begin{aligned}
				&
				\mathcal{Z}=\frac{1}{1-T}=\sum^{\infty}_{n=0}e^{-n{\rm ln}\beta}
				={\rm Tr}e^{-\hat{N}{\rm ln}\beta},\\
				&
				\mathcal{O}=e^{-\hat{N}{\rm ln}\beta},\\
				&
				\hat{N}={\rm diag}[0,1,2,\cdots,\infty],
			\end{aligned}
		\end{equation}
		where we treating the effective temperature $T$ as a complex argument and thus $\mathcal{Z}={\rm Li}_{1}(T)$.
		Here the effective temperature is determined by the partition function such that
		it can be cast into the series form with 
		Then, since $T\partial_{T}{\rm Li}_{1}(T)={\rm Li}_{0}(T)=\frac{T}{1-T}$,
		we obtain 
		\begin{equation} 
			\begin{aligned}
				{\rm Tr}[\mathcal{O}\mathcal{H}_{lo}]=\frac{T^2}{(1-T)^2}
				=\frac{1}{1-(\frac{2}{T}-\frac{1}{T^2})}
				=\sum_{n=0}^{\infty}
				e^{-n{\rm ln}
					(\frac{2}{T}-\frac{1}{T^2})^{-1}
				}
				=\sum_{n=0}^{\infty}E^{m}_{n}
				e^{-n{\rm ln}
					\beta}
				={\rm Tr}e^{-\hat{N}{\rm ln}\beta'},
			\end{aligned}
		\end{equation}
		where
		$E^{m}_{n}
		=e^{-n{\rm ln}
			(2\beta-\beta^2)^{-1}
			+n{\rm ln}\beta}$ is the many-body energy\cite{Lantagne},
		and $\beta'=
		(\frac{2}{T}-\frac{1}{T^2})^{-1}
		$.Thus $\mathcal{H}_{nH}=e^{-\hat{N}{\rm ln}\beta'+\hat{N}{\rm ln}\beta}$,
		and
		$	\frac{{\rm Tr}[\mathcal{O}\mathcal{H}_{lo}]}{\mathcal{Z}}$ in Eq.(\ref{1117})
		is the "thermal" average estimation of the local term $\mathcal{H}_{lo}$
		(not necessarily be Hermitian)
		but for the effective temperature which should be distincted from the case with a real temperature.
		While the linear response expansion in terms of GGE can be determined by
		$e^{-\beta\mathcal{H}_{lo}-\sum_{i}\beta_{i}\mathcal{H}_{lo}'}=e^{-\hat{N}{\rm ln}\beta}$.
		
		Thus we have $\mathcal{S}
		={\rm ln}\mathcal{Z}+\frac{T}{1-T}
		={\rm Li}_{1}(T)+{\rm Li}_{0}(T)$ where
		the first and second terms correspond to the thermal effect of initial state and
		that emergent during the thermal relaxation (the temporal fluctuation satisfying the ETH ansatz),
		and the physical correspondence for the first term can 
		be found by the integral of the Fermi-Dirac distribution.
		And in the zero-temperature limit,
		$\mathcal{S}\sim 2T$, by virtue of the asymptotic behavior
		$\lim_{T\rightarrow 0}\frac{{\rm Li}_{1,2}(T)}{T}=1$,
		while the "thermal" energy (the contribution to non-Hermicity) 
		is $T\mathcal{S}\sim 2T^2$,
		which can be seem from the nearly linear scaling in zero-temperature limit
		as shown in Fig.\ref{sca}.
		
		The robust correlation between two subsystems can be seem from the persistent particle-hole symmetry
		and the absence of sign-dependent asymptotic degeneracy.	
		\begin{equation} 
			\begin{aligned}
				{\rm Tr}[{\bf \Psi}_{k}]
				={\rm Tr}[{\bf \Phi}{\bf J}_{k}{\bf \Phi}^{T}]
				=\sum_{i}{\rm Tr}[|\varepsilon_{i}\rangle\langle\varepsilon_{i}|{\bf J}_{k}]
				=
				\sum_{i}\langle \varepsilon_{i}|{\bf J}_{k}|\varepsilon_{i}\rangle
			\end{aligned}
		\end{equation}
		where ${\bf J}_{k}=\frac{-1}{\beta_{0}}
		{\rm ln}(|\varepsilon_{i}\rangle\langle\varepsilon_{i}|)$
		where $\beta_{0}$ is a complex quantity that can be identified according to the trace class restriction,
		i.e., ${\rm Tr}[|\varepsilon_{i}\rangle\langle\varepsilon_{i}|]
		={\rm Tr}[{\rm exp}[-\beta_{0}{\bf J}_{k}]]=1$, ($\forall i$).
		Also, note that the powered identity matrices
		$|\varepsilon_{i}\rangle\langle\varepsilon_{i}|^{\frac{-1}{\beta_{0}}}$
		are raised to be full-rank such that its logarithm is meanful
		(which is possible for a complex power).
		${\bf \Phi}$ is the symmetry matrix whose rows are the eigenvectors 
		of Hermitian operator $H$ (${\bf \Phi}^{T}={\bf \Phi}^{-1}$).
		Using
		replica trick, we have
		\begin{equation} 
			\begin{aligned}
				{\rm ln}|\varepsilon_{i}\rangle\langle\varepsilon_{i}|^{\frac{-1}{\beta_{0}}}
				=\lim_{\alpha\rightarrow 0}
				\frac{(|\varepsilon_{i}\rangle\langle\varepsilon_{i}|^{\frac{-1}{\beta_{0}}})^{\alpha}-1}
				{\alpha}
				=\lim_{\alpha\rightarrow 0}
				\frac{|\varepsilon_{i}\rangle\langle\varepsilon_{i}|^{\frac{\alpha}{\beta_{0}}}-1}
				{\alpha}.
			\end{aligned}
		\end{equation}
		As the $|\varepsilon_{i}\rangle$ is the orthogonal and normalized basis of Hermitian operator $H$,
		\begin{equation} 
			\begin{aligned}
				{\rm Tr}[{\bf \Psi}_{k}]
				=
				\sum_{i}\langle \varepsilon_{i}|{\bf J}_{k}|\varepsilon_{i}\rangle
				={\rm dim}[H]\lim_{\alpha\rightarrow 0}
				\frac{i\eta(\alpha)}{\alpha},
			\end{aligned}
		\end{equation}
		where $\eta(\alpha)$ is a real infinitely small quantity as a function of $\alpha$.

		Since for such a bipartite system, to guarantees the non-Hermiticity,
		one of the subsystem must be rank-deficient while the other one must be full-ranked.
		Here we always denote the rank-deficient subsystem by ${\bf \Psi}_{1}$
		with the corresponding tensor ${\bf J}_{1}$,
		and the full-ranked subsystem by ${\bf \Psi}_{2}$
		with the corresponding tensor ${\bf J}_{2}$.
		Then the tensor of the subsystem with full-rank can be defined as
		${\bf J}_{2}=\frac{-1}{i\eta\beta}
		{\rm ln}|\varepsilon_{i}\rangle\langle\varepsilon_{i}|^{i\eta}$.
		Our numerical result shows that the valid $\eta$ that meets the trace restriction exhibit no much difference among different eigenvectors.
		As both the ${\bf J}_{2}$ and $({\bf J}_{1}-{\bf J}_{2})$ are full-ranked,
		we can, in perspective of validated thermodynamics, obtain the tensors as $
		{\rm Tr}[|\varepsilon_{i}\rangle\langle\varepsilon_{i}|
		({\bf J}_{1}-{\bf J}_{2})
		=\langle\varepsilon_{i}|
		({\bf J}_{1}-{\bf J}_{2})|\varepsilon_{i}\rangle,
		{\rm Tr}[|\varepsilon_{i}\rangle\langle\varepsilon_{i}|
		{\bf J}_{1}
		=\langle\varepsilon_{i}|
		{\bf J}_{1}|\varepsilon_{i}\rangle$.

		To focus on the non-Hermitian contribution regarding the effect of identically distributed narrow channels,
		we donot consider the effect from electronic transport in momentum space
		by consider the case of nearly flat bands.
		By sampling randomly the Gaussian distributed eigenkets we can obtain a non-Hermitian operator whose
		level distribution follows GOE,
		and the diagonal/off-diagonal elements follow the diagonal/off-diagonal ETH ansatz.
		We further divide the non-Hermioan operator $\mathcal{H}_{nH}$ into two subsystems,
		each of whose diagonal elements are uncorrelated and satisfy the diagonal ETH,
		while the off-diagonal elements are inevitable correlated to guarantees the vanishing long-time fluctuation for the off-diagonal part of $\mathcal{H}_{nH}$.
		Then as the usual treatment to the Kubo formula in calculating the conductivity is
		using the Luttinger-independent-electrons-approximation,
		i.e., $\langle v_{x}(0)v_{y}(\varepsilon)\rangle\approx \langle v_{x}(0)\rangle\langle v_{y}(\varepsilon)\rangle$.
		This treatment ignore part of the uncommutative effects.
		In our system,
		$\mathcal{H}_{nH}$ is consist of two trace-one subsystems,
		whose eigenvalues can find a correspondence to the diagonal elements,
		thus we define the effective velocities
		$v_{eff}^{x(y)}=\frac{\langle\varepsilon_{i}|{\bf  J}_{1(2)}|\varepsilon_{j}\rangle}
		{\langle\varepsilon_{i}|\varepsilon_{j}\rangle}$
		while the inner product between eigenvectors of Hermitian operator $\mathcal{H}$ plays the role 
		of effective momentum which always equals to one due to the orthonormality
		(spatial effect from momentum space are thus removed such that the result only depends on the size of Hilbert space).

		\begin{figure}
			\centering
			\includegraphics[width=0.4\linewidth]{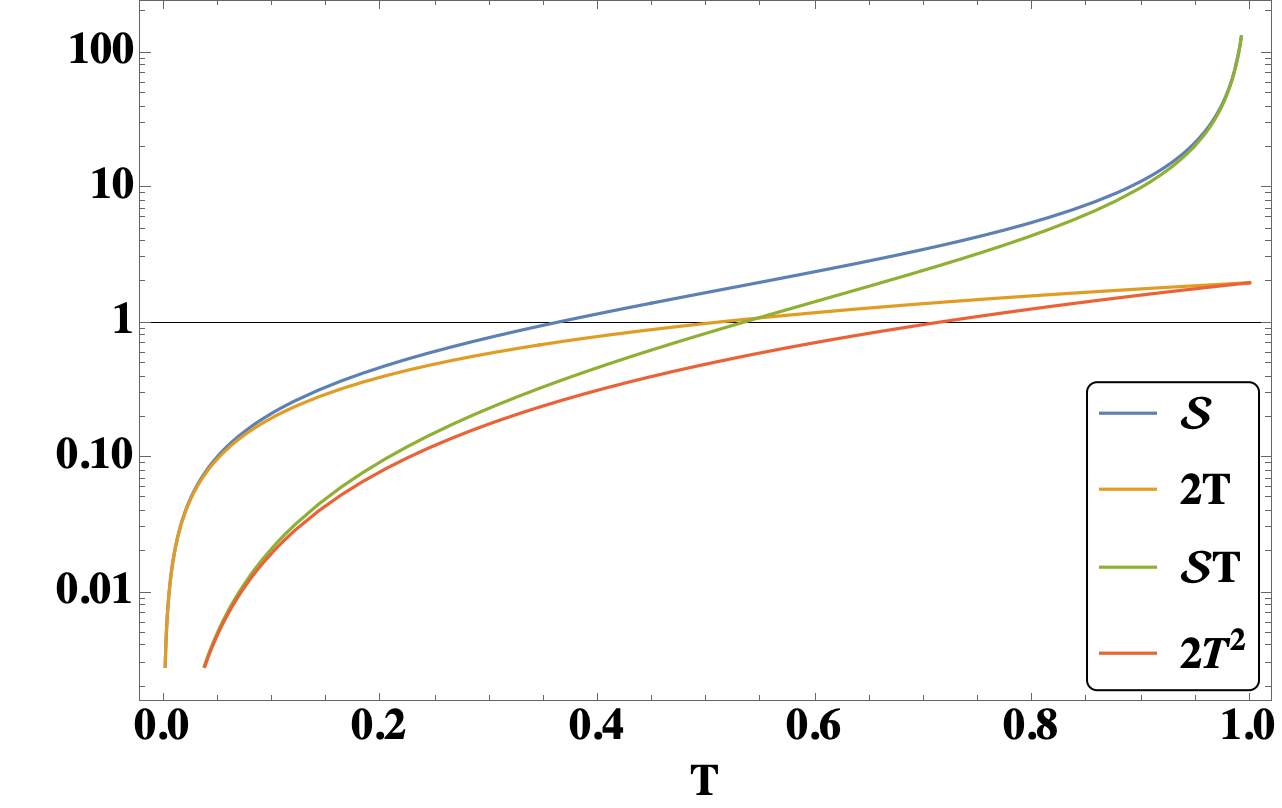}
			\caption{Thermal entropy $\mathcal{S}$ and related thermal energy $\mathcal{S}T$.
			}
			\label{sca}
		\end{figure}

	\end{small}
\end{document}